# Charging dynamics of dopants in helium nanoplasmas

Andreas Heidenreich

*Kimika Fakultatea, Euskal Herriko Unibertsitatea (UPV/EHU) and Donostia International Physics Center (DIPC), P.K. 1072, 20080 Donostia, Spain, and IKERBASQUE, Basque Foundation for Science, 48011 Bilbao, Spain*

Barbara Grüner

*Physikalisches Institut, Universität Freiburg, 79104 Freiburg, Germany*

Dominik Schomas

*Physikalisches Institut, Universität Freiburg, 79104 Freiburg, Germany*

Frank Stienkemeier

*Physikalisches Institut, Universität Freiburg, 79104 Freiburg, Germany*

Siva Rama Krishnan

*Department of Physics, Indian Institute of Technology - Madras, Chennai 600036, India*

Marcel Mudrich*

*Physikalisches Institut, Universität Freiburg, 79104 Freiburg, Germany*

* Corresponding author: mudrich@physik.uni-freiburg.de

# Charging dynamics of dopants in helium nanoplasmas

We present a combined experimental and theoretical study of the charging dynamics of helium nanodroplets doped with atoms of different species and irradiated by intense near-infrared (NIR) laser pulses ($\leq 10^{15}$ Wcm$^{-2}$). In particular, we elucidate the interplay of dopant ionization inducing the ignition of a helium nanoplasma, and the charging of the dopant atoms driven by the ionized helium host. Most efficient nanoplasma ignition and charging is found when doping helium droplets with xenon atoms, in which case high charge states both of helium (He$^{2+}$) and of xenon (Xe$^{21+}$) are detected. In contrast, only low charge states of helium and dopants are measured when doping with potassium and calcium atoms. Classical molecular dynamics simulations which include focal averaging generally reproduce the experimental results and provide detailed insights into the correlated charging dynamics of guest and host clusters.

## 1. Introduction

Nanoplasmas formed in the intense near-infrared irradiation of neutral nanoscale particles ($\approx$1-100 nm) have continued to draw considerable attention due to the intriguing aspects of their dynamics ensuing in them such as the formation of an ionization avalanche on the few-femtosecond timescale [1-5]. For near-infrared (NIR) laser pulses with photon energies below atomic ionization potentials of the atomic constituents of these nanoclusters, ionization is initiated by nonlinear field photoionization of the atoms [6-7]. The crucial role of generating initial seed electrons which trigger the ignition of a nanoplasma has been experimentally directly demonstrated. By either irradiating pristine nanoclusters by additional weak extreme ultraviolet pulses [3], or by inserting dopant atoms with low ionization threshold intensity into the host clusters [4,8], the threshold intensity for the NIR laser field to ignite a nanoplasma is lowered by up to two orders of magnitude. For the extreme case of helium (He) nanodroplets the addition of a small number of dopant atoms (e. g. xenon, Xe) inside the droplets was shown to be sufficient to trigger complete ionization giving rise to high yields of detected He$^+$ and He$^{2+}$ ions even at intensities where the pristine He droplet remains unaffected by the laser pulse [4,5].

Efficient charging of the cluster occurs due to the driving of the seed electrons by the NIR laser field to acquire average kinetic energies in large multiples of the ponderomotive energy. This energy is expended in inelastic collisions within the cluster leading to an avalanche of electron impact ionization of the constituent atoms and ions [6,7]. Upon subsequent expansion of the nanoplasma, resonant coupling with the driving laser field can occur when the plasmon frequency sweeps over the laser frequency as the electron density falls off. Resonant light absorption then further promotes the charging and heating of the expanding nanoplasma. As a consequence, electron-ion recombination is suppressed, giving rise to increased final charge states of detected ions.

Although a generic understanding of the intense NIR pulse dynamics of single component clusters from few-femtosecond to picosecond timescales has been achieved [6,7], the ignition of the nanoplasma as well as the charging of dopants embedded inside ionizing host clusters remains to be investigated in greater detail [9,10]. A series of experiments involving dopant clusters grown in He nanodroplets have led to similar conclusions in accordance with the generic nanoplasma dynamics of single component clusters [2,11] However, in certain cases the presence of a He shell surrounding the dopant cluster was found to distinctly influence the charging dynamics of the dopant atoms [12]. In a comparative study of lead (Pb) clusters formed in He droplets and as free clusters which were irradiated by two time-delayed pulses, the maximum yield of high charge states was reached at significantly earlier delay times for Pb clusters embedded in He droplets [13]. This was rationalized in later studies by the fast expansion of the ionized He shell around the embedded dopant cluster leading to a shifting of the plasmon resonance conditions in the expanding nanoplasma to earlier times [14]. For silver atoms embedded in He nanodroplets the transient reduction of the final charge states was attributed to charge transfer occurring between multiply charged dopant atoms and the surrounding He [15]. In contrast, in the case of Xe clusters embedded in He droplets the final Xe charge states were increased with respect to bare Xe clusters of the same size when irradiated by NIR pulses stretching over the He plasmon resonance [2]. This was attributed to the additional charging effect induced by the He component of the nanoplasma.

In a recent study we have elucidated the roles of the electronic structure as well as the location of dopants inside (rare-gas) or on the surface (alkali, alkaline earth metals) of the host He droplet in triggering avalanche ionization at peak laser intensities $10^{13}$ – $10^{15}$ W/cm$^2$ [8]. Both the ability of dopant atoms to provide seed electrons as well as their location in close contact with the He droplet were found to be crucial parameters. Counterintuitively, those dopants with the lowest ionization energies (potassium, K) turned out to be the least efficient ones in igniting avalanche ionization of the He droplets, whereas high yields of He$^+$ and He$^{2+}$ ions were measured when doping the droplets with Xe. This was rationalized by Xe unifying several favorable properties for acting as a seed for He nanoplasma ignition: Its ability to easily donate more than one seed electron per dopant atom, its location inside the droplets, as well as its low heat of cluster formation which limits shrinkage of the droplets due to evaporation of He atoms. The purpose of the present work is to investigate the interrelation between ignition of a He nanoplasma yielding large He ion signals due to the presence of the dopant acting as seeds, and the charging up of dopant atoms to high charge states. To this end, we present experimental mass-over-charge spectra for the same dopant species as in our previous work (Xe, calcium (Ca), K), accompanied by systematic numerical simulations of the same systems.

## 2. Experiment

The experimental setup has been described in detail elsewhere [8,14]. In short, a beam of doped He nanodroplets is irradiated with amplified femtosecond near-infrared (NIR) laser pulses (center wave length $\lambda = 800$ nm, pulse length $t_{FWHM} = 220$ fs) of variable intensity up to about $I = 1 \times 10^{15}$ W/cm$^2$. The peak intensity is determined from comparing appearance intensities of various charge states of rare gas atoms in the background gas with the literature [16]. We estimate an uncertainty of this value of about a factor two. Ion mass spectra are recorded using a standard time-of-flight mass spectrometer. From the recorded mass-over-charge spectra we subtract the background spectra measured when blocking the He droplet beam so as to suppress contributions from ionized background gas.

A beam of He nanodroplets is generated in a continuous expansion of pressurized He ($p_0 = 50$ bar) into vacuum out of a cold nozzle ($T_0 = 18$ K) with a diameter of 5 μm. At these expansion conditions, the mean droplet size is $\langle N \rangle \approx 5000$ atoms [17] before droplet shrinkage upon doping. Doping of the droplets with small clusters of either rare gas or metal atoms is achieved by passing the droplets through a doping cell which contains atomic vapor at adjustable pressure. In the case of doping with Xe, the vapor pressure is adjusted by leaking Xe gas into the doping chamber using a dosing valve. In the case of doping with the alkali metal K or with the alkaline earth metal Ca, the dopant vapor pressure is adjusted by controlling the temperature of the doping cell which contains a piece of the elementary dopant material. When doping the droplets with more than one dopant atom by increasing the dopant vapor pressure above the level of single atom pick-up, the dopant atoms aggregate to form small clusters located in the bulk of the droplets (Xe) or at the droplet surface (K, Ca) [18]. Under these doping conditions, the He droplets undergo scattering due to the transfer of transverse momentum, as well as droplet shrinkage due to evaporation of He atoms induced by the deposition of kinetic and binding energy (dopant-dopant and dopant-He). Both effects cause a depletion of the droplet-correlated ion and electron count rates. The average number of dopants attached to the He droplets is inferred from the measured dopant vapor pressure in combination with numerical simulations of the pick-up process [19].

3. Simulations

The molecular dynamics (MD) simulation method for the interaction of a cluster with the electric and magnetic field of a linearly polarized NIR Gaussian laser pulse was described previously [8,20,21]. All atoms and nanoplasma electrons are treated classically, starting with a cluster of neutral atoms. Electrons enter the MD simulation, when the criterion for tunnel ionization (TI), classical barrier suppression ionization (BSI) or electron impact ionization (EII) is met. This is checked at each atom at every MD time step, using the local electric field at the atoms as the sum of the external laser electric field and the contributions from all other ions and electrons of the cluster. Instantaneous TI probabilities are calculated by the Ammosov-Delone-Krainov formula [22], EII cross sections by the Lotz formula [23] taking the ionization energy with respect to the atomic Coulomb barrier in the cluster [24]. The effect of chemical bonding on the valence shell ionization energies of K and Ca dopants is disregarded.

Coulomb potentials between ions, and smoothed Coulomb potentials for ion-electron and electron-electron interactions are used. Interactions involving neutral atoms are disregarded except for a Pauli repulsive potential between electrons and neutral He atoms. The Pauli repulsive potential is expressed as a sum of pairwise forth-order Gaussian functions centered at every neutral He atom, $V(r_{ij}) = V_0 \exp(-r_{ij}^4/4\sigma^4)$, with $r_{ij}$ being the He-electron distance, $V_0 = 1.1$ eV [25], and the exponent $\sigma = 1.2$ Å is chosen such that the effective range of $V(r_{ij})$ is about half of the average He-He distance (3.6 Å) in the neutral He droplet.

The binding potentials of $He_2^+$ and of other $He_n^+$ complexes are not implemented, so that the MD simulations cannot account for $He_2^+$ formation explicitly; we can only estimate an upper bound of the $He_2^+$ abundance from the remaining ground state neutral He atoms and $He^+$ ions at the end of each trajectory. Neutral He atoms and $He^+$ ions which are formed by three-body electron-ion recombination (TBR) are Rydberg state atoms and are therefore excluded from the estimate of the $He_2^+$ production. Electron-ion pairs which are found within a cutoff distance of 2 Å at the end of each trajectory (temporal length 0.7-1.8 ps) are taken to be recombined and the ion charge state abundances are corrected accordingly.

He ion and dopant ion signals are laser-intensity averaged over the three-dimensional focus volume [26] in the range $1.45 \times 10^{12}$ - $5 \times 10^{15}$ Wcm$^{-2}$. Due to the high sensitivity of the droplet evolution to initial conditions, the results are averaged over sets of 5 to 100 trajectories per doped droplet and laser intensity. Moreover, surface-doped droplets (K and Ca) are averaged over their parallel and perpendicular orientations of the dopant-droplet axis with respect to the laser polarization axis unless mentioned explicitly (notation for doping the droplet interior will be C, and X and Y for surface doping in parallel and perpendicular orientation of the dopant-$He_N$ complex with respect to the polarization axis, respectively). The temporal width of the Gaussian pulse intensity envelope is $t_{FWHM} = 200$ fs, slightly lower than in the experiment (220 fs). As stated in section 2, the experimental average He droplet size is $\langle N \rangle \approx 5000$ atoms where a considerable droplet shrinkage by He evaporation takes place upon doping, so that the majority of doped droplets are smaller. In our MD simulations we chose $He_{459}$ and $He_{2171}$ as samples of the droplet size distribution.

For the He droplets we assume a fcc structure with an interatomic distance of 3.6 Å [27]. The dopant clusters are assembled according to the principle of densest packing of tetrahedra and to form, as far as possible, spherical shapes. We use the following interatomic distances: K-K 4.56 Å (taken as the average interatomic distance in a $K_{20}$ cluster) [28], Ca-Ca 3.9 Å (average value for Ca clusters) [29], Xe-Xe 4.33 Å (bulk), He-Xe 4.15 Å [30], He-K 7.13 Å [31], He-Ca 5.9 Å [32]. In case of surface doping with Ca, we assume a dimple depth of 7 Å [33]. According to Ancilotto et al. [34], a single K atom is located in a dimple of depth 2.3 Å. Since such a shallow dimple cannot be implemented in a fcc lattice of discrete He atoms, we neglect the dimple for K dopants.

## 4. Results and discussion
### *4.1. Simulated ionization dynamics*

To describe the microscopic processes evolving in the course of the droplet ionization, Fig. 1 depicts several time-dependent properties obtained from a single trajectory example of a $He_{2171}$ droplet doped with a $Xe_8$ cluster sitting in the droplet center (C) at $I = 10^{14}$ Wcm$^{-2}$. Shown are the normalized electric field envelope in panel (a), the average Xe charge $\langle q_{Xe} \rangle$ in panel (b), the average He charge $\langle q_{He} \rangle$, the number $n_p$ of nanoplasma electrons per atom and the radius $R$ of the expanding He droplet as multiples of the initial radius $R_0$ in panel (c). $R$ is taken as the distance of the outermost He atom from the droplet center-of-mass.

Ionization starts with TI or BSI at a single dopant atom in the rising edge of the laser pulse at $t_{dop}$ = -87 fs in this single-trajectory example, immediately followed by further TI, BSI and EII of the dopant. In this trajectory, the average Xe charge, panel (b), rises to 1.6 elementary charges $e$ before the first He atom is ionized. This occurs at $t_{He-1}$ = -83 fs, marked by a red vertical line in Fig. 1. He ionization is always induced by EII; the role of the dopant is to provide the seed electrons and to assist EII by lowering the Coulomb barrier at He by the field of the dopant cations. Somewhat later, EII of the He droplet becomes avalanche-like, manifested by a steep rise of the average He charge in panel (c). The beginning of the avalanche ionization of the He droplet we term 'ignition' [8], marked by a green vertical line $t_{ignit}$ = -71 fs. We define an average He threshold charge of 0.1 as an empirical criterion for the detection of ignition; the choice of the exact value is uncritical in view of the rapid charging process. EII is by far the dominating ionization channel (typically > 95% of all ionizations).

We term the delay between the first dopant ionization and ignition 'incubation time' [8]. During this time, which lasts 16 fs in this trajectory, EII of He competes with a partial drain (outer ionization [35]) of the seed electrons. In general, this competition is not always in favor of EII. It turns out that, depending on slight variations of the trajectories' initial conditions but for the same pulse parameters, either He ionization does not take place at all, ceases after a few He atoms, or ignition occurs. The sensitivity to the initial conditions is high for those dopant sizes and pulse parameters for which the occurrence of ignition is on the knife's edge. Consequently, for a quantitative comparison with experiment one has to average over multiple trajectories. The ignition probability is then given as the fraction of the trajectories in which ignition takes place. In case of the $He_{2171}Xe_8(C)$ droplet, the ignition probability is 1.

Fig. 1d) exhibits the laser energy $W_{abs}$ absorbed by the nanoplasma, the total kinetic energy $T_e$ of all nanoplasma electrons and the energy $E_{EII}$ expended on EII. $W_{abs}$ is the integral of the power absorption $P$,

$$W_{abs}(t) = \int_{-\infty}^{t} P(t') dt', \text{ where } P(t) = \sum_i \left( e q_i \vec{v}_i(t) \cdot \vec{\varepsilon}(t) \right). \quad (1)$$

$q_i$ and $\vec{v}_i$ are the charge in units of $e$ and the velocity vector of particle i, respectively. $\vec{\varepsilon}$ is the laser electric field. $P(t)$, shown in panel (e), is almost exclusively given by the contribution of the nanoplasma electrons. As in previous works of Peltz et al. [36] and of Krishnan et al. [14], $W_{abs}(t)$ rises steeply during the droplet avalanche ionization and converges thereafter. The power absorption continues to show strong oscillations which fade away with decreasing laser electric field strength, but do not lead to further net energy absorption. Such pronounced energy absorptions were assigned phenomenologically to the sweeping over a resonance of the nanoplasma electron density [14,36]. The total electron kinetic energy $T_e$ passes its maximum shortly before $W_{abs}(t)$ converges to its final value.

The disadvantage of $W_{abs}(t)$ as an indicator for resonance is that its value is largely dependent on the laser electric field and on the strongly varying number of electrons $n(t)$ because of avalanche EII and outer ionization. To reveal the characteristics of resonance, which may be identified with a positive cycle-averaged scalar product of electron velocity and driving force, we divide $P(t)$ by $n(t)$ and the absolute value of the laser electric field, $\varepsilon(t) = |\vec{\varepsilon}(t)|$,

$$\xi(t) \equiv \frac{P(t)}{n(t)\varepsilon(t)}. \qquad (2)$$

One could call $\xi(t)$ a reduced power absorption per electron. The integral

$$\Xi(t) = \int_{-\infty}^{t} \xi(t')dt' \qquad (3)$$

is then a measure for the net reduced energy absorption. Fig. 1f) shows the functions $\xi$ and $\Xi$ for the single-trajectory example for $He_{2171}Xe_8$ as black and blue lines, respectively. $\xi(t)$ is an oscillating function with two maxima of the envelope. Together with positive $\Xi(t)$ values, this suggests that the nanoplasma electrons pass through two resonances. The first resonance starts immediately after the dopant ionization and ends with ignition. The second resonance begins towards the termination $t_{term}$ (marked by a blue vertical line in Fig. 1) of the nanoplasma formation, cf. $\langle q_{Xe} \rangle$, when also the droplet radius has expanded by a factor of two, panel 1c). It is known [6,7,37] that droplet expansion crucially influences the plasmon resonance condition of the nanoplasma due to the changing electron density. The first resonance is quickly crossed over as the number of nanoplasma electrons steeply increases around $t = t_{ignit}$ without notable expansion of the droplet. Only after the droplet has sufficiently expanded towards the end of nanoplasma formation, the electron density has dropped down so as to match the resonance condition again. About 3/4 of the energy absorption takes place in the second resonance. Note that during the period of reduced $\xi(t)$ between the resonances, $\langle q_{Xe} \rangle$ (panel 1b)) increases at a reduced rate and subsequently takes a final jump, once the second resonance is reached. In the second resonance also outer ionization becomes substantial as manifested by the decrease of $n_p$ in panel 1c). Pristine Xe clusters and He droplets for $I \geq 5 \times 10^{14}$ Wcm$^{-2}$ (i.e., for intensities when TI rates become notable) show the same behavior: A first resonance as long as only a limited number of atoms is ionized, followed by a period in which the $\xi$ function assumes small values, and a second resonance during which most of the energy absorption takes place.

Saalmann et al. have analyzed the nanoplasma oscillations in terms of a driven damped harmonic oscillator [6,37]. Here we refrain from presenting such a complete analysis. Instead we discuss the phase $\phi(t)$ [Fig. 1f] of the nanoplasma oscillation with respect to the driving force given by the laser electric field. Starting with $\phi \approx \pi$ at $t = t_{dop}$, during the first resonance $t_{dop} \leq t \leq t_{ignit}$, $\phi$ rapidly crosses over the value $\pi/2$. During this initial charging period it drops down to $\pi/5$ which is consistent with the common notion of the plasmon frequency rising above the laser frequency as a consequence of the build-up of a high electron density. Subsequently, $\phi(t)$ again reaches the value of $\pi/2$ and finally levels out to $\phi = \pi$ at long expansion times as expected for a low electron density bound to a shallow potential well created by the expanded droplet.

The observation of Döppner et al. [13] that embedding of a dopant cluster (in their case a Pb cluster of up to 120 atoms) in a He droplet leads to an earlier resonance as compared to the bare dopant cluster, is not born out by our MD simulations of small embedded Xe dopant clusters because Xe is by about 1/3 lighter than Pb and the smaller clusters considered in this work expand faster and reach their resonance condition earlier [38]. We believe that a shift of the resonance to earlier times would be observed for larger Xe dopant clusters as well. The final average Xe charge in the $He_{2171} \cdot Xe_8$ droplet is much higher than in the free $Xe_8$ dopant and almost as high as in the pristine $Xe_{2171}$ cluster. In this single-trajectory example of $He_{2171} \cdot Xe_8$ we obtain $\langle q_{Xe} \rangle = 9.4$. The trajectory-set average for $I = 10^{14}$ Wcm$^{-2}$ is $\langle q_{Xe} \rangle = 10.1$. For comparison, with a pristine cluster $Xe_8$ only $\langle q_{Xe} \rangle = 4.1$ is reached; in contrast, the value for $Xe_{2171}$ is $\langle q_{Xe} \rangle = 11.3$.

Fig. 2 shows a single-trajectory example of a doped $He_{2171}$ droplet at low pulse peak intensity, $I = 10^{13}$ Wcm$^{-2}$. Since Xe dopants require a minimum pulse peak intensity of $I \approx 5 \times 10^{13}$ Wcm$^{-2}$, we chose Ca as a dopant in this example. Thus, a $Ca_{23}$ cluster is positioned in a dimple on the droplet surface such that the dopant–droplet axis is parallel with respect to the laser polarization axis (X-direction). The long-term averaged He charge is only $\langle q_{He} \rangle = 1.2$, the average Ca charge is $\langle q_{Ca} \rangle = 2.2$.

The striking difference of the time evolution of Fig. 2 compared to Fig. 1 is the shift of the characteristic instants $t_{ignit}$ and $t_{term}$ to later times, 30 and 116 fs, respectively, compared to -71 and -16 fs for the $Xe_8$ dopant at $I = 10^{14}$ Wcm$^{-2}$ (see the green and blue vertical lines in Fig. 1 and 2). The $\xi$ and $\Xi$ functions [Fig. 2f)] indicate the presence of a first resonance at times $t_{He-1} \leq t \leq t_{ignit}$, but the second resonance for times $t > t_{term}$ is missing. The droplet expands more slowly than at $I = 10^{14}$ Wcm$^{-2}$ [cf. Figs. 1c) and 2c)] and the energy consumption $E_{EII}$ takes a much larger part of the absorbed laser energy $W_{abs}$ [panel (d)], causing a strong damping of the driven nanoplasma electron oscillations. Although the phase $\phi$ assumes values $\approx \pi/2$ in the time interval $t_{ignit} \leq t \leq t_{term}$, $\xi$ does not show large-amplitude oscillations and $\Xi$ does not increase steeply, as one would expect for a resonance. The reason might be the strong damping. For a smaller droplet, $He_{459} \cdot Ca_{23}(X)$, where less atoms are ionized, droplet expansion and near termination of EII take place several tens of femtoseconds earlier, and the second resonance is observed even at pulse peak intensities as low as $I = 10^{13}$ Wcm$^{-2}$ (not shown).

### 4.2. Analysis of the charging process

In this section we show that the ignition time of a droplet is crucial for its charging process. As indicated in section 4.1., in a set of trajectories with different initial conditions, the times $t_{dop}$ and $t_{He-1}$ at which the first dopant and He atom are ionized, respectively, as well as the ignition times $t_{ignit}$ form distributions (even if the ignition probability is unity). Fig. 3 shows two examples of such distributions for the clusters $He_{2171} \cdot Xe_{13}(C)$ and $He_{2171} \cdot Ca_{13}(X)$ at $I = 5 \times 10^{13}$ Wcm$^{-2}$ obtained from bundles of 100 trajectories. While for $He_{2171} \cdot Xe_{13}$ the three distributions overlap, they are well separated for $He_{2171} \cdot Ca_{13}$, shifted to earlier times within the laser pulse envelope and also show much longer incubation times than for Xe doping. In general, $t_{dop}$, $t_{He-1}$ and $t_{ignit}$ depend on the dopant element, dopant cluster size and pulse peak intensity. The shift of the distributions to earlier times for Ca (for the same pulse peak intensity) can be attributed to its lower first ionization energy.

To illustrate this dependence, we show in Fig. 4 the average ignition times in panel a), the ignition probabilities in panel b), and the final average He charge (without the charge reducing effect of TBR) in panel c) as a function of the dopant cluster size for $He_{2171}$ droplets doped with $K_n$, $Ca_n$ and $Xe_n$ at $I = 5 \times 10^{13}$ Wcm$^{-2}$. The following trends become apparent:

(1) With increasing the dopant cluster size, ignition is shifted to earlier times.
(2) For the same number of dopant atoms, ignition is shifted to later times in the order Xe > K > Ca. That is, Xe tends to ignition occurring at latest times.
(3) For surface states (K and Ca), parallel orientation (X) of the droplet in the laser field leads to earlier ignition than perpendicular orientation (Y), although the effect is masked here to a large extent by the statistical sampling error of the relatively small trajectory sets.
(4) The average He charge $\langle q_{He}\rangle$ increases with increasing dopant cluster size.

The results show that although Xe can induce ignition with a smaller number of dopant atoms and is from this point of view the most efficient dopant element among those considered here, K, Ca and Xe, it features the latest ignition times. This is a consequence of its higher first ionization energy (12.1 eV) compared to K and Ca (4.3 and 6.1 eV, respectively), which causes the first Xe ionization to be delayed towards the maximum of the pulse peak envelope. Points (1) and (4) together imply that the final average He charge is very sensitive to the ignition time.

How close the correlation between $t_{ignit}$ and $\langle q_{He}\rangle$ is becomes apparent when both quantities are plotted against each other. Fig. 5 shows correlation diagrams of the single-trajectory average He charge $\langle q_{He}\rangle$ (without averaging over the trajectory set but averaged over all He atoms of the droplet and without the charge reduction effect of TBR) vs. $t_{ignit}$. a) He$_{2171}$ and b) He$_{459}$ are considered to assess the influence of droplet sizes. Included are MD simulation results for the three pulse peak intensities $I = 10^{13}$, $2 \times 10^{13}$ and $5 \times 10^{13}$ Wcm$^{-2}$ for K, Ca, and Xe in surface and interior dopant positions. Every data point represents one trajectory. Fig. 5 shows that irrespective of the dopant element, dopant cluster size and dopant location, the data points lie, with relatively little deviations, on curves which are well determined by the droplet size and pulse parameters. In contrast, no correlations are found between the pairs of quantities ($t_{dop}$, $t_{He-1}$), ($t_{dop}$, $t_{ignit}$) and ($t_{He-1}$, $t_{ignit}$).

As expected, small values of $t_{ignit}$ generally result in the most efficient charging of the He droplets for all dopant species. With decreasing values of $t_{ignit}$, the curves become horizontal where $\langle q_{He}\rangle$ reaches its saturation value $\langle q_{He}\rangle_{sat}$ which is determined by the droplet size and by the pulse parameters. In the range 50 fs $\leq t_{ignit} \leq$ 150 fs the curves drop sharply, as at late times the droplet cannot absorb a sufficient amount of laser energy. Although the data points are not distinguished by the dopant sizes, larger dopants constitute the majority of data points at low ignition times, as the preceding discussion of Fig. 4 implies. At $I = 10^{14}$ Wcm$^{-2}$ (not shown in Fig. 5) almost all data points lie on the horizontal part of the curve at $\langle q_{He}\rangle_{sat} \approx 2$. That is, when ignition occurs at $I = 10^{14}$ Wcm$^{-2}$, it almost always leads to a complete ionization of the droplet, irrespective of the ignition instant. He$_{459}$ as the smaller droplet leads to lower average ion charges, as known from pristine rare-gas clusters [20]. Notice that the curves for $I = 10^{13}$ and $2 \times 10^{13}$ Wcm$^{-2}$ do not contain data points of Xe doped droplets, since TI of Xe requires intensities $I \geq 5 \times 10^{13}$ Wcm$^{-2}$.

The central issue of this work is to assess the interrelation between final dopant and He charges. Fig. 6 exhibits six examples for correlation diagrams of final average dopant charges vs. final average He charges without TBR, where every data point represents a trajectory. The first three examples, K and Ca doped droplets at $I = 2 \times 10^{13}$ Wcm$^{-2}$ [panels (a) and (b)] and Xe doped droplets at $I = 5 \times 10^{13}$ Wcm$^{-2}$ [panel (c)] show the most pronounced correlations. With some deviations, the data points follow well-separated curves, distinguished by droplet size, dopant location [surface or interior (C)] and, for surface doping, also by the orientation [parallel (X) or perpendicular(Y)] of the doped droplet in the laser field. In the examples of panels (a-c), most of the data points are located on parts of the curve which have a finite positive slope, indicating a clear dependence of the dopant charge on the He charge and vice versa. As the formation of the He nanoplasma proceeds efficiently, the dopants concurrently charge up to high charge states due to a high density of driven electrons which resonantly couple to the light field, assisted by lowering of the Coulomb barriers by the electric field of the nanoplasma ions.

The correlation is much less pronounced for Ca doped droplets at $I = 10^{13}$ Wcm$^{-2}$, see panel 6d). A considerable number of data points lie on the horizontal part of the curve, where the dopant charge is insensitive to the nanoplasma conditions characterized by $\langle q_{He} \rangle$ values. This is due to the high threshold energy of 50.9 eV that has to be overcome for ionizing Ca$^{2+}$ to create Ca$^{3+}$. For Ca dopants at $I = 5 \times 10^{13}$ Wcm$^{-2}$ [panel (e)], and for Xe dopants at $I = 10^{14}$ Wcm$^{-2}$ [panel (f)], almost all data points fall into vertically spread distributions, either at $\langle q_{He} \rangle \approx 0$ or at the saturation value $\langle q_{He} \rangle = \langle q_{He} \rangle_{sat}$ of the respective droplet size. The vertical branch near the maximum He charge corresponds to early ignition times which can push the dopant charges to higher values even when $\langle q_{He} \rangle$ reached its saturation value. The corresponding $\langle q_{He} \rangle$ data points are found in the horizontal distribution of the data for $t_{ignit} < 0$ in Fig. 5. From the presented examples we conclude that two conditions must be met for dopant and He charges to show strong correlations:

(1) The average He charge must be sensitive to the ignition time. This condition is realized in the range of low and intermediate intensities $I = 10^{13} - 5 \times 10^{13}$ Wcm$^{-2}$ where dopant-induced ignition of the He droplet sets in but He ionization remains incomplete.

(2) For the given peak intensity, various higher dopant charge states must be accessible to EII. This condition is well fulfilled for Xe where the lowest ionization energies are rather closely spaced (12.1, 21.0, 31.1 eV), whereas it is less well fulfilled for K which has a low first ionization energy energy (4.3 eV) but the second ionization energy is much higher (31.6 eV).

The discussion of the charging of dopants and He droplets and their mutual interactions so far was focused on short-time or inner ionization dynamics. However, to link the simulations with experimental results, charge recombination occurring in the course of plasma expansion at longer times needs to be taken into account. To this end, we elucidate the effect of charge recombination on both dopant and He charge states. Figs. 7 exhibits the extent of TBR at He for the pulse peak intensities $I = 10^{13}$ a), $2 \times 10^{13}$ b), and $5 \times 10^{13}$ Wcm$^{-2}$ c). In each of the three correlation diagrams, single-trajectory averaged He charges with TBR are plotted against the corresponding values without TBR. Accordingly, data points lying in the lower triangles of the graphs indicate TBR to be active, whereas data points on the diagonal indicate that TBR is absent for the corresponding trajectory.

Irrespective of the dopant species and dopant cluster size, all data points form only two branches determined by the two considered droplet sizes, He$_{459}$ and He$_{2171}$. TBR is very pronounced for data points $\langle q_{He} \rangle < 1.2$. These data points correspond to late ignition times (cf. Fig. 5) and consequently to a low energy absorption of the droplet. Smaller droplet sizes generally lead to less TBR [7]. At $I = 10^{13}$ Wcm$^{-2}$, TBR is very pronounced but decreases with increasing peak intensity. At $I = 5 \times 10^{13}$ Wcm$^{-2}$, TBR is already negligible for K and Ca doping, but still can be found to some extent for Xe doping, as manifested by the predominance of black and grey symbols in the lower triangle of panel c). The more pronounced occurrence of TBR for Xe doping is connected with the later ignition times (c.f. Fig. 5) which facilitate TBR by the somewhat lower laser energy absorption. At $I = 10^{14}$ Wcm$^{-2}$, TBR does not play a role for this pulse duration (not shown in Fig. 7).

Correlation diagrams in which dopant charges with and without TBR are plotted against each other are shown in Fig. 8. The dopant charges without TBR exhibit the following trends:

(1) For $I \leq 2 \times 10^{13}$ Wcm$^{-2}$, Ca leads to higher average charges than K, for $I = 5 \times 10^{13}$ Wcm$^{-2}$ the hierarchy Xe > Ca > K is exhibited. The average charges at $I = 5 \times 10^{13}$ Wcm$^{-2}$ are up to around 5 for K, 5.5 and 7 for surface and interior doping with Ca, respectively, and up to 10.5 for Xe dopants in He$_{2171}$ droplets.
(2) The dopant charges increase with increasing He droplet size, in accord with results for pristine rare-gas clusters which show the general trend of increasing ion charges with increasing cluster size [7,20].
(3) For Ca dopants for which simulation results are available for all the three dopant locations / orientations, the trend C > X > Y is observed which reflects the degree of overlap of the driven electron cloud inducing EII of the dopants atoms.

From Fig. 8 we infer the following trends in TBR:

(4) As in the case of He, TBR is abundant for K and Ca dopants at $I \leq 2 \times 10^{13}$ Wcm$^{-2}$. At $I = 5 \times 10^{13}$ Wcm$^{-2}$, TBR is found for Xe dopants to some extent.
(5) The data points of surface and interior doping states form different branches in the diagrams, as it becomes apparent for Ca doping at $I = 10^{13}$ and $2 \times 10^{13}$ Wcm$^{-2}$ [panels (a) and (b)] for which simulation results are available for both doping locations. Interior doping leads to more recombination than surface doping due to the presence of a high electron density in the center of a cold nanoplasma.
(6) For the lowest intensity ($I = 10^{13}$ Wcm$^{-2}$), the highest ion charge states are most susceptible for TBR. For higher intensities intermediate charge states are mostly affected. This is the result of two opposite effects. On the one hand, higher ion charges favor recombination. On the other hand, the generation of higher charge states is accompanied by massive energy absorption which suppresses TBR.

### 4.3. Comparison experiment-simulation

The results of the MD simulation show that the mutual interaction of the Xe dopant cluster with the He droplet leads to high charge states of both Xe and He provided that ignition of a nanoplasma takes place. In contrast, for He droplets doped with small K and Ca clusters nanoplasma ignition is less likely and mostly low dopant charge states are created for the same conditions. This result is indeed confirmed by experiments, where we measure time-of-flight ion mass-over-charge spectra for He nanodroplets doped with the species Xe, Ca, and K for various sizes of the dopant clusters. Typical spectra recorded at a laser peak intensity $I = 10^{15}$ Wcm$^{-2}$ are depicted in Fig. 9. The indicated average numbers of dopant atoms per He droplet are obtained from the measured dopant vapor pressures [19].

The spectra of Xe-doped He droplets [Fig. 9a) and b)] are dominated by the He ion signals He$^+$, He$^{2+}$, and He$_2^+$. In the absence of Xe-doping, these signals are barely visible on the scale used in Fig. 9. Aside from He ions, the mass spectra reveal Xe$^+$ and Xe$_n^+$ cluster ions up to $n = 3$ (not shown) which may be limited by the acceptance of our mass spectrometer. In addition, an extended series of multiply charged Xe$^{k+}$ ions is observed up to the charge state Xe$^{21+}$ [see inset of Fig. 9a)]. These high charge states of Xe are only seen for Xe embedded in He nanodroplets; with atomic Xe gas at the same laser intensity we measure quickly decreasing relative yields of higher charge states in the order $Xe^+ : Xe^{2+} : Xe^{3+} : Xe^{4+} : Xe^{5+}$ given by 1 : 0.3 : 0.15 : 0.05 : 0.005. The measured mass-over-charge spectra are analyzed by performing nonlinear fits of the $Xe^{k+}$ ion series using a sum of shifted Lorentzian functions with variable amplitudes. The result of such a fit is shown in Fig. 9b) as a red line.

In the mass-over-charge spectra of He nanodroplets doped with Ca and K shown in Fig. 9c) and d), the absolute intensities of the He ion signals fall below those for Xe doping for any doping level; for Ca doping, the maximum $He^+$ intensity reached at about 3 Ca dopant atoms amounts to 8% of that achieved with Xe doping and for K doping the $He^+$ intensity remains below 0.3% of that of Xe doping [8]. Thus, the mass-over-charge spectra for Ca and K doping are dominated by ions of the dopant atoms and of small dopant clusters as well as by progressions of charged complexes with a number of He atoms $[CaHe_m]^+$, $[CaHe_m]^{2+}$, and $[KHe_m]^+$. Highest signals are given by partly and fully valence shell-ionized $Ca^+$, $Ca^{2+}$ and $K^+$, where $Ca^+$ and $Ca^{2+}$ occur with nearly equal abundance for all doping levels. At the highest possible dopant vapor pressures above which massive beam destruction sets in, the higher charge states $K^{2+}$ and $Ca^{3+}$ appear but remain weaker by factor 40 than the corresponding valence ionization signals. $Ca_n^+$ clusters are detected up to $n=20$ and doubly charged clusters $Ca_n^{2+}$ up to $n = 19$. For K, only singly charged clusters $K_n^+$ are measured up to $n = 21$. Stable "snowball" complexes $[CaHe_m]^+$, $[CaHe_m]^{2+}$, and $[KHe_m]^+$ have been observed previously using various ionization schemes and for $m$ reaching up to the size of the intact He droplet [12,39-41]. Direct ionization of the atomic vapor effusing out of the doping cell yields $Ca^+$ and $Ca^{2+}$ as well as $K^+$ ions but no cluster ions.

A compilation of the most representative ion signals in dependence of the vapor pressure of dopants as measured in the experiment is shown in Fig. 10. When doping with K [panel a)], the maximum count rate of the by far most abundant $K^+$ signal is reached at a K partial pressure inside the 1 cm long doping cell of $2.5 \times 10^{-4}$ mbar. This signal stems from direct ionization of K atoms and small $K_n$ clusters without significant influence of the He droplets in the ionization process. The much weaker signals of $He^+$ and $K^{2+}$ are reached at a factor $\approx 10$ higher doping pressure, at which $K_n$ clusters are formed and nanoplasma ignition occurs to some extent. The doping dependence of $Ca^+$ and $Ca^{2+}$ ions [panel b)], mostly from direct dopant ionization, closely follows the one of $K^+$. Low yields of $Ca^{3+}$ from nanoplasma ignition already appear at slightly lower dopant partial pressure ($\approx 5 \times 10^{-4}$ mbar) than the $K^{2+}$ signal in the K case. When doping with Xe [panel c)], high yields of $He^+$ and $He^{2+}$ around $8 \times 10^{-4}$ mbar $\times$ cm and comparatively much lower yields of high charge states of Xe ions peaking at slightly higher doping pressures indicate highly efficient nanoplasma ignition [8]. The fact that the low $Xe^{k+}$ charges, $k<10$, show elevated yields even at high doping pressures $> 4 \times 10^{-3}$ mbar $\times$ cm we attribute to the formation of pure Xe clusters by complete evaporation of the He off the droplets in the course of massive doping and Xe cluster aggregation.

We note that (i) the regimes of ionization – direct multiphoton or strong-field ionization of the dopants and dopant-induced ignition of a nanoplasma – can be controlled by varying the doping pressure. In the K case the two regimes are well separated due to the extreme ratio of first and second ionization energies, whereas in the Ca and Xe cases there is a smooth transition from one regime to the other. (ii) in all three cases the maximum yield of He ions is reached at lower doping pressures than the signal maxima of dopant ions stemming from the nanoplasma. This is mostly due to shrinking of the droplets setting in when dopant clusters form upon multiple doping. Under such conditions the proportion of He vs. dopant atoms in the mixed clusters shifts toward larger dopant contributions. Massive destruction of the droplets at high doping pressures evidently reduces both He and dopant ion signals.

The observation of high dopant charge states for the case of Xe doping, for which also high yields of He ions are measured, confirms the close connection between He avalanche ionization and dopant charging found in the MD simulations. In an attempt to directly compare the results of the simulation with the experiment, we have computed dopant charge distributions for selected dopant cluster sizes. A complete MD simulation of the ion signals requires (i) averaging over all intensities which contribute in the focal volume to the ion signal, (ii) averaging over the size distribution of doped droplets, (iii) averaging over the dopant size distribution, (iv) for surface dopant locations, averaging over the orientations of the dopant-droplet axis relative to the laser polarization, as well as (v) averaging over multiple trajectories with different initial conditions. For the simulated dopant charge distributions, Fig. 11, we have carried out only points (i), (iv) and (v). The minimum pulse peak intensity for focal averaging was taken as the BSI threshold intensities for K and Ca, $I = 1.45 \times 10^{12}$ and $5.6 \times 10^{12}$ Wcm$^{-2}$, respectively, and $I = 5 \times 10^{13}$ Wcm$^{-2}$ for Xe, at which TI becomes important. To elucidate the dependence of the dopant charge distributions on the He droplet size, we derive distributions for He$_{459}$ and He$_{2171}$ droplets as well as for the bare dopant clusters, shown in Fig. 11.

While in the case of doping He$_{2171}$ with small Xe clusters we obtain dopant charges peaked around Xe$^{10+}$, dopant charges for Ca and K clusters remain rather low. Already when doping He droplets with one single Xe atom, the distribution displays contributions of strongly enhanced Xe charge states compared to the free Xe dopant clusters. The charge distribution for $n_{Xe} = 1$ is bimodal; the part at low charges corresponds to unignited droplets and therefore resembles the distribution of the bare dopant atom. For Xe the dopant charge distributions largely depend on the He droplet size and for $n_{Xe} < 8$ on the dopant cluster size. The latter reflects the rapidly increasing ignition probability with increasing the dopant cluster size. For K and Ca the situation is different for two reasons. First, K and Ca are more inert towards inner shell ionizations. For impact energies $\leq 500$ eV, Xe has 18 accessible charge states whereas K and Ca have only 9 and 10, respectively. As it became apparent in Figs. 6 and 8, the average dopant charges without TBR at $I = 5 \times 10^{13}$ Wcm$^{-2}$ are up to 5 and 5.5 for surface doping of K and Ca, while Xe in its interior doping site reaches a value of 10.5. Second, in focal averaging the outweighing contribution to the K and Ca charge state distribution originates from pulse peak intensities $I < 5 \times 10^{13}$ Wcm$^{-2}$. 99 and 93% of the focal volume of K and Ca, respectively, are located in this low-intensity periphery, where valence states of K and Ca are ionized but where ignition requires larger dopant clusters. Ca is always doubly ionized; Ca$^+$ abundances are the result of recombination. In contrast, already small Xe clusters can ignite He droplets at the threshold intensity $I = 5 \times 10^{13}$ Wcm$^{-2}$ where Xe itself is ionized.

For the cases of doping with K and Ca the agreement between the MD simulation and experiment is good in that almost exclusively valence shell-ionized atoms are present for the doping levels set in the experiment. The experimentally observed cluster ions cannot be reproduced by the MD simulation because interactions between charge and neutral particles are completely neglected. As for the charge spectra of Xe, we measure much more extended distributions reaching up to higher charge states than in the MD simulation. For one part the Lotz formula for EII cross sections takes into account only direct EII whereas it is known [24] that the excitation-autoionization and recombination-autoionization channels can drastically exceed the contribution of direct EII. For the other part the lack of higher Xe charge states is due to the missing averaging over broad distributions of both Xe dopant cluster sizes $n_{Xe}$ and He droplet sizes $N$, present in the experiment, whereas fixed sizes are set in the MD simulation. The results of the MD simulation for the two different He droplet sizes $N = 459$ and 2171 show indeed substantial shift of the charge state distribution. Thus, a simulation taking into account averaging over the extended distribution of $N$ would certainly yield a broad distribution of $\langle q_{He} \rangle$ closer to the measured one. Such a simulation would require a tremendous numerical effort, though, from which we desist at this stage.

For the sake of presenting a direct comparison between experiment and MD simulation we compute the average values of dopant and He charge states and plot them in Fig. 12 as a function of the number of dopant atoms in or on the droplets (left and right columns, respectively). In general the experimental data are in reasonable agreement with the MD simulation; in contrast to the stark increase of the yield of ions when doping He droplets with small numbers of dopant atoms [4,8] the average dopant charges show only little variation with the dopant number for K and Ca as well as for Xe in the range $n_{Xe} > 14$. In the MD simulation $\langle q_{He} \rangle \approx 2$ for all shown dopant species and clusters sizes. In the experiment, we mostly measure higher yields of $He^+$ than of $He^{2+}$ resulting in $\langle q_{He} \rangle \leq 1.5$. In particular when doping with Ca we measure a reduction of $\langle q_{He} \rangle$ with rising average dopant numbers which reflects the presence of $He_2^+$ and $He_3^+$ ions in the spectra. This deviation from the simulation is attributed to additional processes which are not included in the MD simulation: Partial neutralization of ions by charge transfer in the incompletely ionized nanoplasma [12], and formation of dimer and trimer He ions. Presumably the effect of averaging over the focus volume and droplet sizes tends to smear out the He charge distribution towards the lower charge state $He^+$ more than what is accounted for in the MD simulation.

Although the simulated value $\langle q_{He} \rangle \approx 2$ shows no variation with the number of Xe dopant cluster size, which speaks for complete ionization of the He droplet, $\langle q_{Xe} \rangle$ continuously rises as the Xe cluster size increases from 1 to 10. This behavior is reflected by the vertical branch in the correlation diagram $\langle q_{Xe} \rangle$ vs. $\langle q_{He} \rangle$ at $\langle q_{He} \rangle \approx 2$ in Fig. 6f) and, as discussed in section 4.2. The experimental values of $\langle q_{Xe} \rangle$ can only be reliably determined for Xe dopant numbers >14 [see Fig. 9a)]. The $\langle q_{Xe} \rangle$ values range between 11 and 14, slightly higher than in the MD simulation. Note that the determination of $\langle q_{Xe} \rangle$ by performing a non-linear fit of experimental data brings about a large uncertainty of $\langle q_{Xe} \rangle$ which we estimate to ±3. This is due to the complexity of the model function used for fitting the data which takes into account the large background signal from $He^+$ and $He_2^+$. The drop of $\langle q_{Xe} \rangle$ towards large dopant numbers is a consequence of the destruction of the doped droplet beam. The average charge states of Ca and K, of $\langle q_{Ca} \rangle \approx 1.4$ and of $\langle q_K \rangle \approx 1$ are in good agreement with the results of the MD simulation.

## 5. Conclusions

In this work we assess the interrelation between dopant-induced ignition of a He nanoplasma and the charging up of dopant atoms to high charge states. The results of experiments, in which He droplets doped with Xe, Ca, or K atoms are ionized by intense NIR pulses, are compared with systematic MD simulations including focal averaging of the simulated ion signals. We find in the experiment that doping of He droplets with Xe atoms results in high charge states of the dopants up to $Xe^{21+}$, accompanied by high yields of $He^+$ and $He^{2+}$ ions. In contrast, only low charge states up to $K^{2+}$ and $Ca^{3+}$ are found for K and Ca. Our MD simulations also show that the formation of a He nanoplasma leads to an enhancement of the dopant charge states and that the different behavior of K and Ca compared to Xe has mainly two reasons. First, the very low first ionization energies of K and Ca correspond to low laser intensity thresholds for field ionization. For these low intensities, which occupy the largest part of the focal volume, ignition of the He nanodroplets would require larger dopants $K_n$ and $Ca_n$ in the size range $n = 20$-$30$ which are not present under the current experimental conditions. Consequently, the dopant ion signals are dominated by $K^+$, $Ca^+$ and $Ca^{2+}$. In contrast, Xe induces ignition as soon as the intensity exceeds the threshold value for Xe field ionization such that ion signals are dominated by high charge states generated in the nanoplasma. Second, our MD simulations show that even when ignition occurs, higher charge states are more easily accessible by EII for Xe than for K and Ca due to the differing spacing between ionization energies.

Furthermore, our MD simulations show a close correlation between the instant of ignition and the final average He charge $\langle q_{He} \rangle$ of the droplet. For a given fixed He droplet size and laser pulse parameters, $\langle q_{He} \rangle$ is determined only by the ignition time, irrespective of the nature of the dopant element, dopant cluster size, dopant location in the droplet. With decreasing ignition time, $\langle q_{He} \rangle$ convergences to a saturation limit $\langle q_{He} \rangle_{sat}$ which depends on the droplet size and pulse parameters. A correlation between the average final dopant charge $\langle q_{dop} \rangle$ and $\langle q_{He} \rangle$ is found for $\langle q_{He} \rangle < \langle q_{He} \rangle_{sat}$, that is for incomplete droplet ionization. For $\langle q_{He} \rangle \approx \langle q_{He} \rangle_{sat}$, $\langle q_{dop} \rangle$ becomes independent of $\langle q_{He} \rangle$ but $\langle q_{dop} \rangle$ is further boosted by the He nanoplasma via EII as the instant of ignition shifts to earlier times with increasing the laser intensity. Three-body ion-electron recombination occurs in abundance for K and Ca doping at low intensities $I \leq 2 \times 10^{13}$ Wcm$^{-2}$. At $I = 5 \times 10^{13}$ Wcm$^{-2}$ recombination occurs to some extent for Xe doping. Recombination reduces the final charge states of both He and dopants.

In principle one could think of exploiting the correlation between long-time nanoplasma properties and ignition time for a simplified approximate MD simulation scheme. Since the long-time properties of the nanoplasma (He charges, nanoplasma electron population, amount of He ion-electron recombination) are in a good approximation independent of the nature of the dopant, it would suffice to simulate a set of long trajectories only for one dopant. Then for any other dopant a set of short trajectories to determine the distribution of ignition times would be sufficient.

*Acknowledgements*

This work is supported by the Deutsche Forschungsgemeinschaft in the frame of the Priority Programme 1840 'Quantum Dynamics in Tailored Intense Fields'. The dissertation of B. Grüner is supported by the Landesgraduiertenförderungsgesetz of Baden-Württemberg. We are grateful to Ivan Infante for prepublication information on inner-shell ionization energies of K and Ca. The authors thank for financial support from the Spanish Ministerio de Economia y

Competividad (ref. no. CTQ2015-67660-P), as well as for computational and manpower support provided by IZO-SGI SG Iker of UPV/EHU and European funding (EDRF and ESF). The participation of S R Krishnan is partially supported by the DST Max Planck India Partner group scheme.

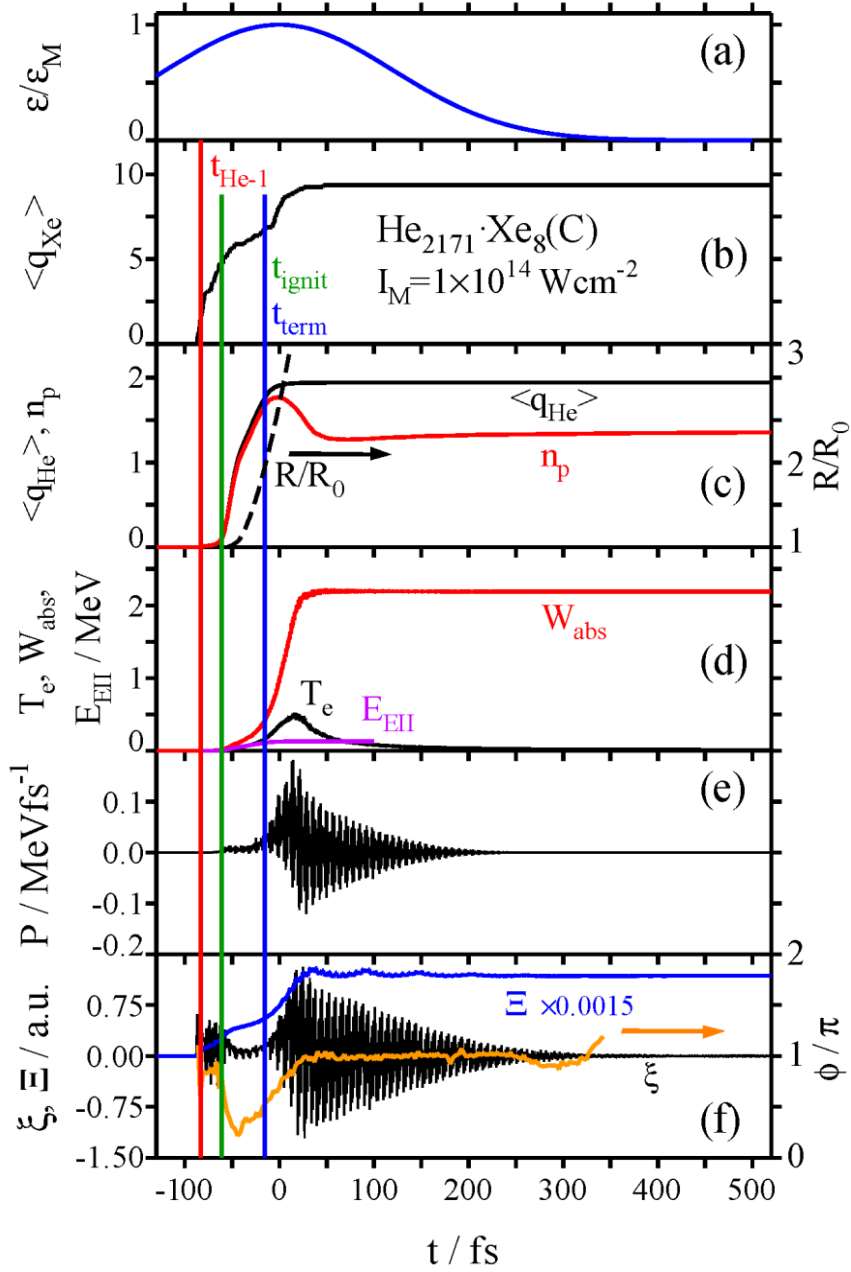

Figure 1. Simulated ionization dynamics of a $He_{2171}$ droplet doped with a $Xe_8$ cluster located in the droplet center for the fixed pulse peak intensity $I = 10^{14}$ Wcm$^{-2}$. (a) Normalized electric field envelope; (b) Average Xe charge; (c) Average He charge, average number $n_p$ of nanoplasma electrons per atom, cluster radius $R/R_0$; (d) Absorbed laser energy $W_{abs}$, electron kinetic energy $T_e$, energy $E_{EII}$ consumed by EII; (e) Absorbed power; (f) Absorbed power ξ per electron and divided by the absolute laser electric field; Ξ is ξ integrated over time to show that net absorption takes place; phase $\varphi/\pi$ between the elongation of the electron cloud center-of-mass and the driving force of the laser electric field. The instants $t_{He-1}$ of the ionization of the first He atom, the ignition time $t_{ignit}$ and the time $t_{term}$ of the near termination of the nanoplasma formation are marked as red, green and blue vertical lines, respectively. $t_{term}$ is taken as the instant when the number of inner ionizations of the droplet has reached 90% of its final value.

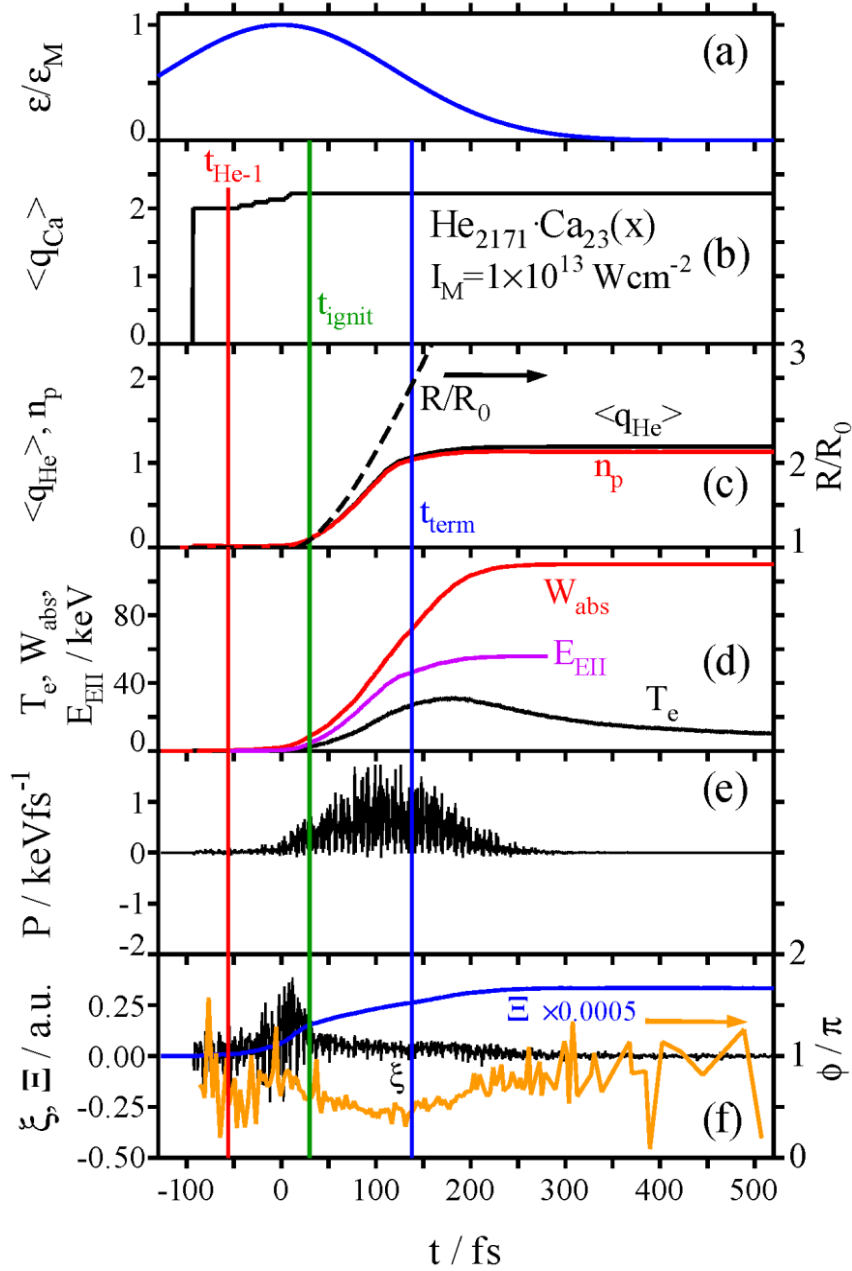

Figure 2. Simulated ionization dynamics of a He$_{2171}$ droplet doped with a Ca$_{23}$ cluster located on the He droplet surface such that the dopant–droplet axis is parallel with respect to the laser polarization axis (X-direction). The pulse peak intensity is fixed to $I = 10^{13}$ Wcm$^{-2}$.

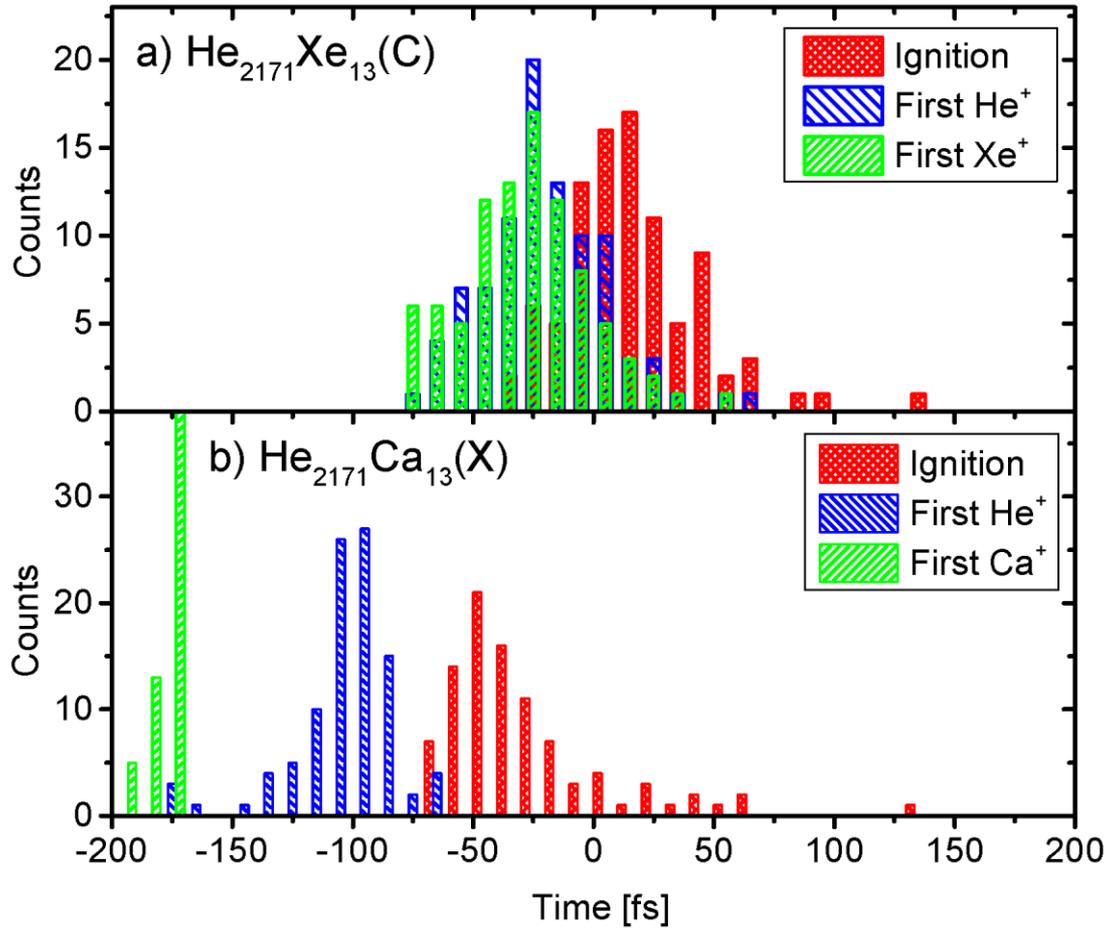

Figure 3. Histograms of (i) the time when the first dopant ionization occurs, $t_{dop}$; (ii) the time of ionization of the first He atom, $t_{He-1}$; (iii) the ignition time $t_{ignit}$, for two selected doped droplets, a) $He_{2171} \cdot Xe_{13}(C)$ and b) $He_{2171} \cdot Ca_{13}(X)$, at $I = 5 \times 10^{13}$ Wcm$^{-2}$.

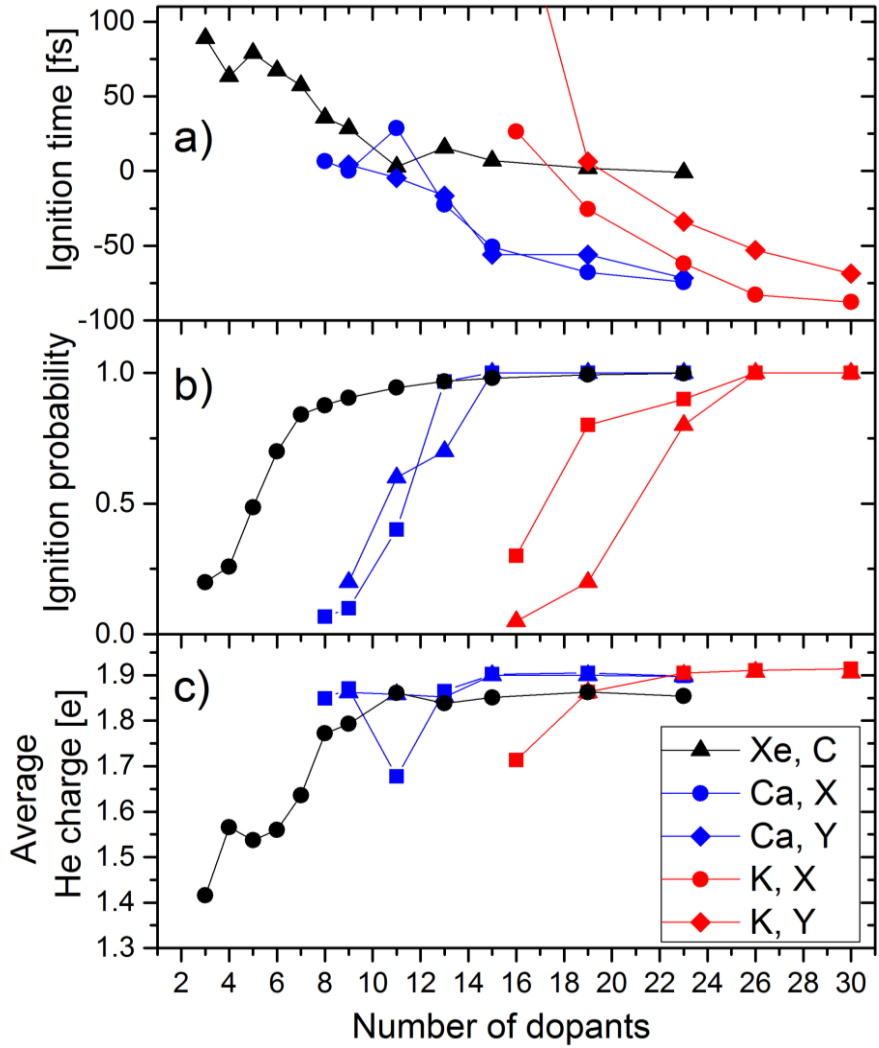

Figure 4. Dependence of various characteristics of nanoplasma ignition on the dopant cluster size dependence. The He charges in panel c) are without TBR but averaged over all He atoms of the droplet and over the trajectory set. C denotes interior doping, X and Y surface doping with the dopant-droplet axis parallel and perpendicular to the laser polarization axis, respectively.

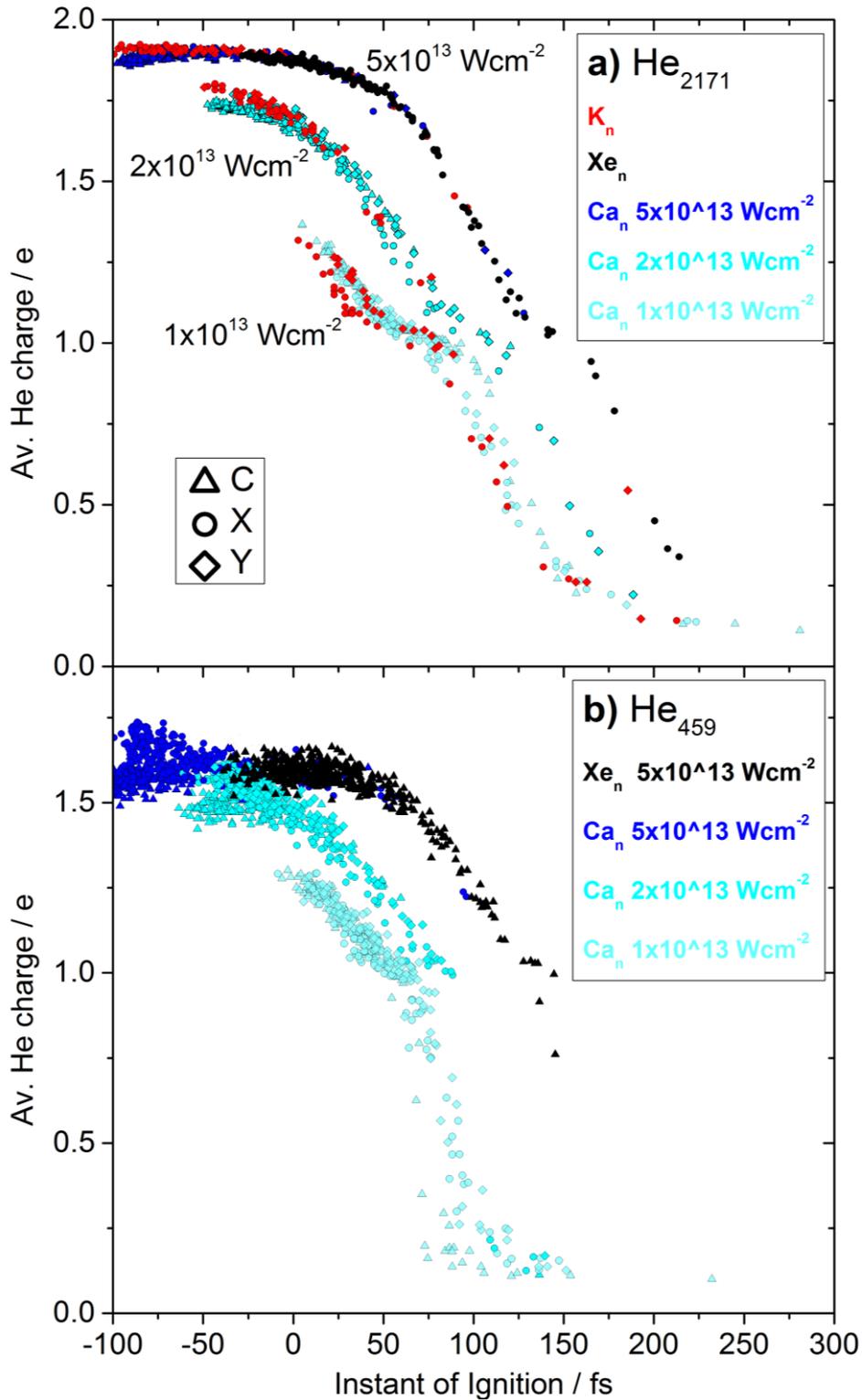

Figure 5. The correlation between the average He charge $\langle q_{He} \rangle$ and the ignition time $t_{ignit}$ for two He droplet sizes and various dopants. The data points are (single-trajectory results) not distinguished by the dopant cluster sizes which are $K_{19-30}$ and $Ca_{11-23}$ for $I = 10^{13}$ Wcm$^{-2}$, $K_{19-30}$ and $Ca_{9-23}$ for $I = 2\times10^{13}$ Wcm$^{-2}$ and $K_{16-30}$, $Ca_{8-23}$ and $Xe_{2-23}$ for $I = 5\times10^{13}$ Wcm$^{-2}$.

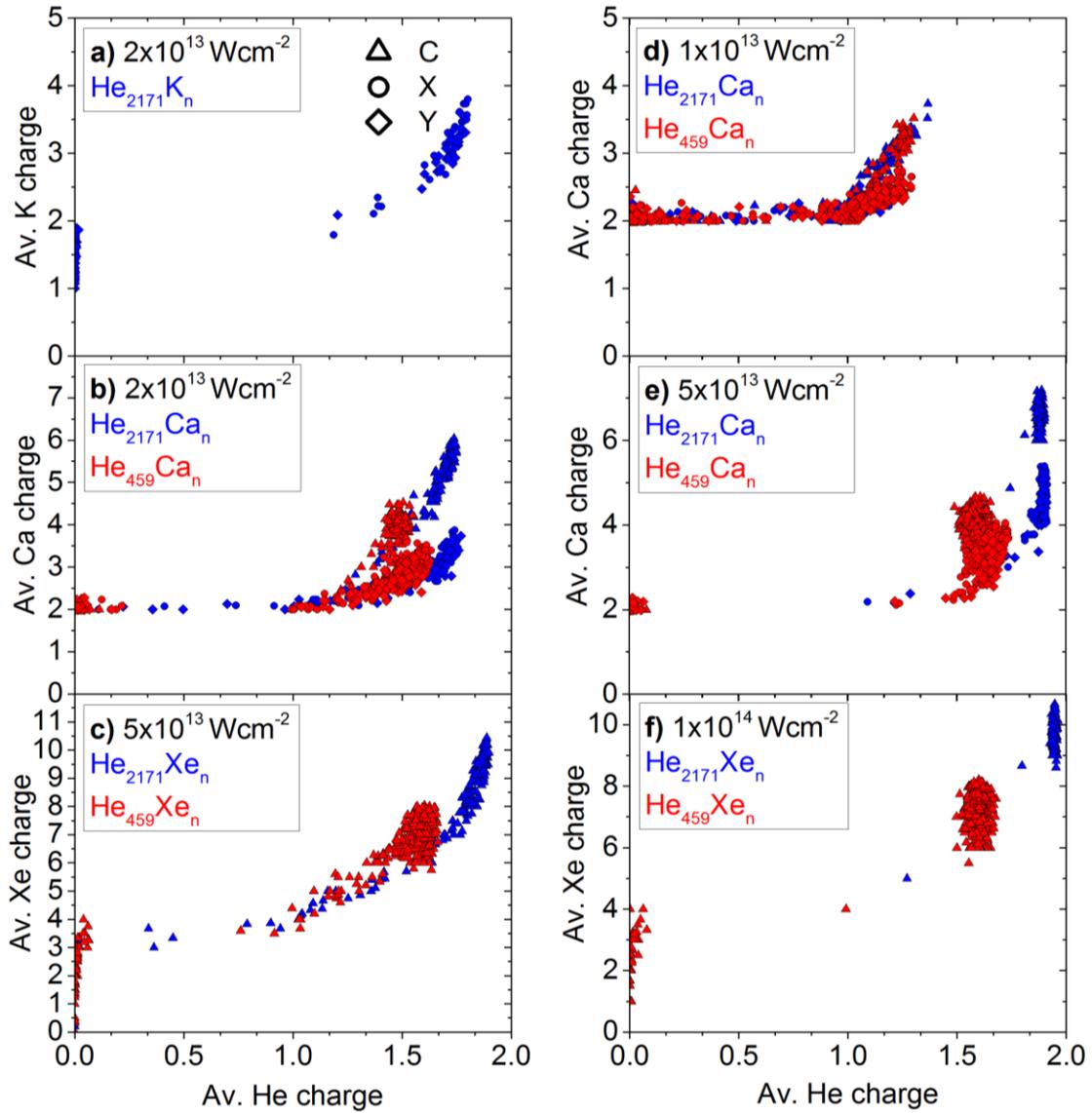

Figure 6. Correlation diagrams of the average dopant charge $\langle q_{\text{dop}} \rangle$ vs. the average He charge $\langle q_{\text{He}} \rangle$ for given dopant species, pulse peak intensities, and droplet sizes. Dopant cluster sizes are not distinguished in this representation.

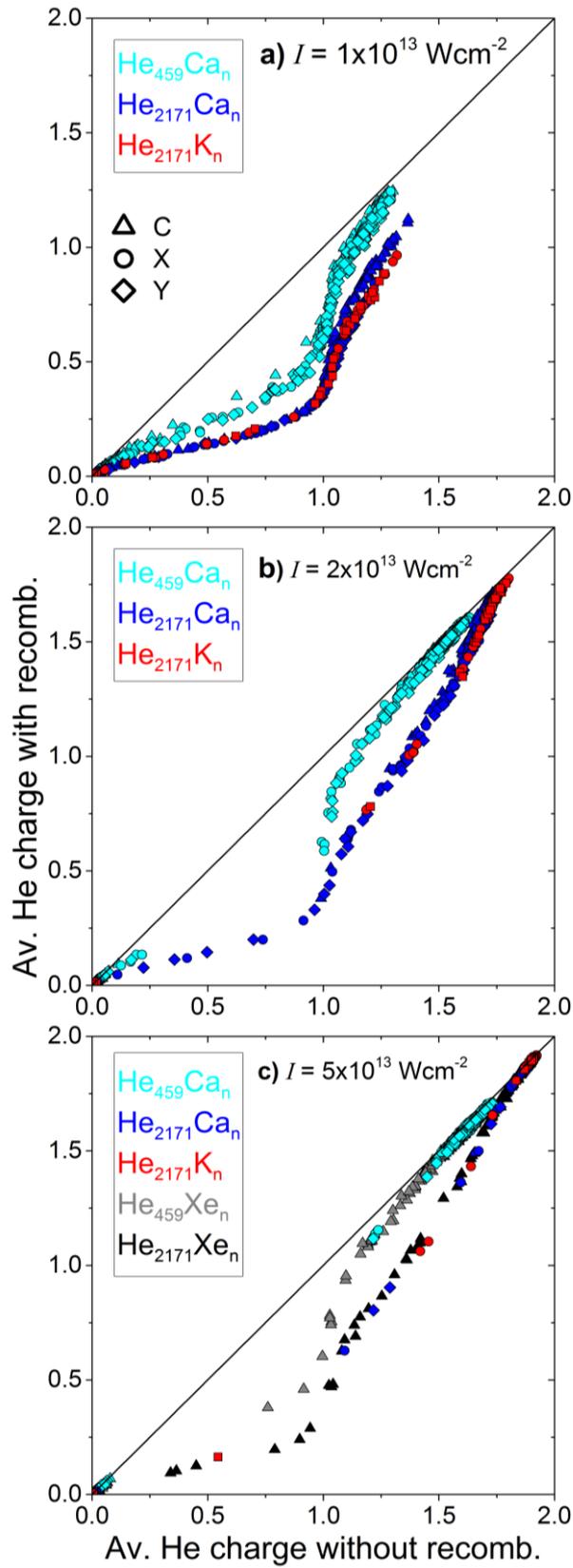

Figure 7. The effect of electron-ion recombination on the average He charges. Dopant cluster sized are not distinguished.

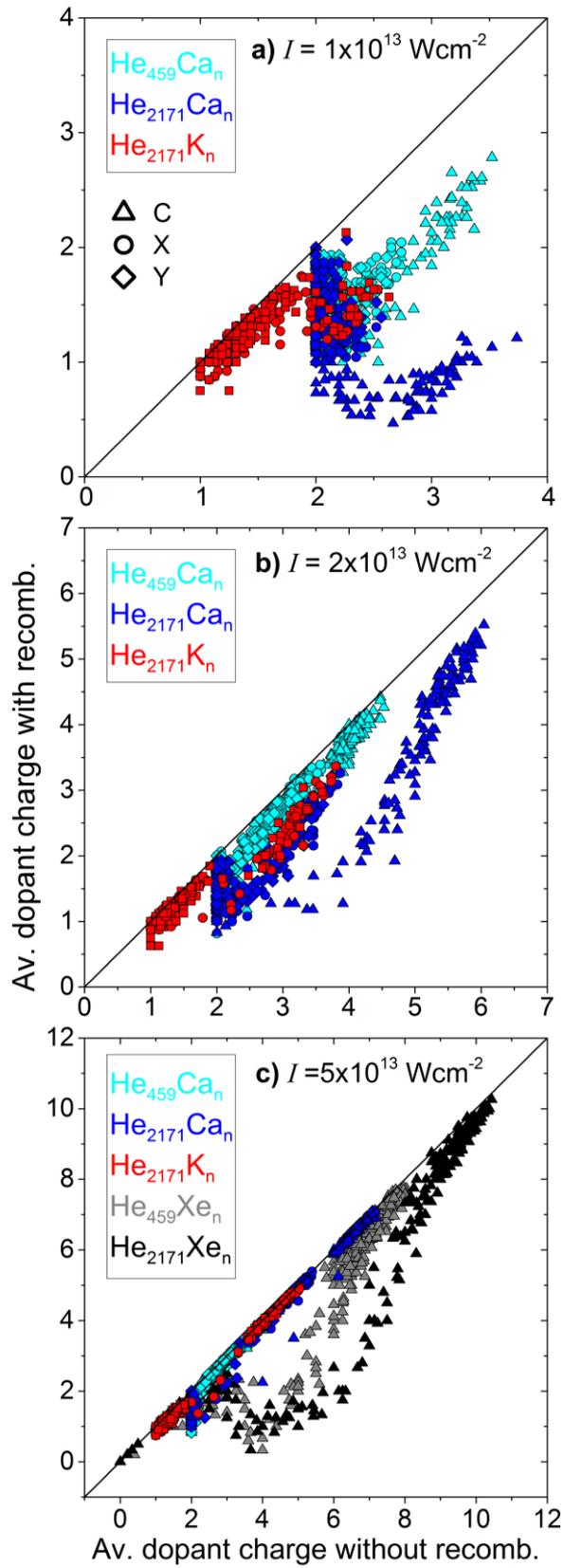

Figure 8. The effect of electron-ion recombination on the average dopant charge states.

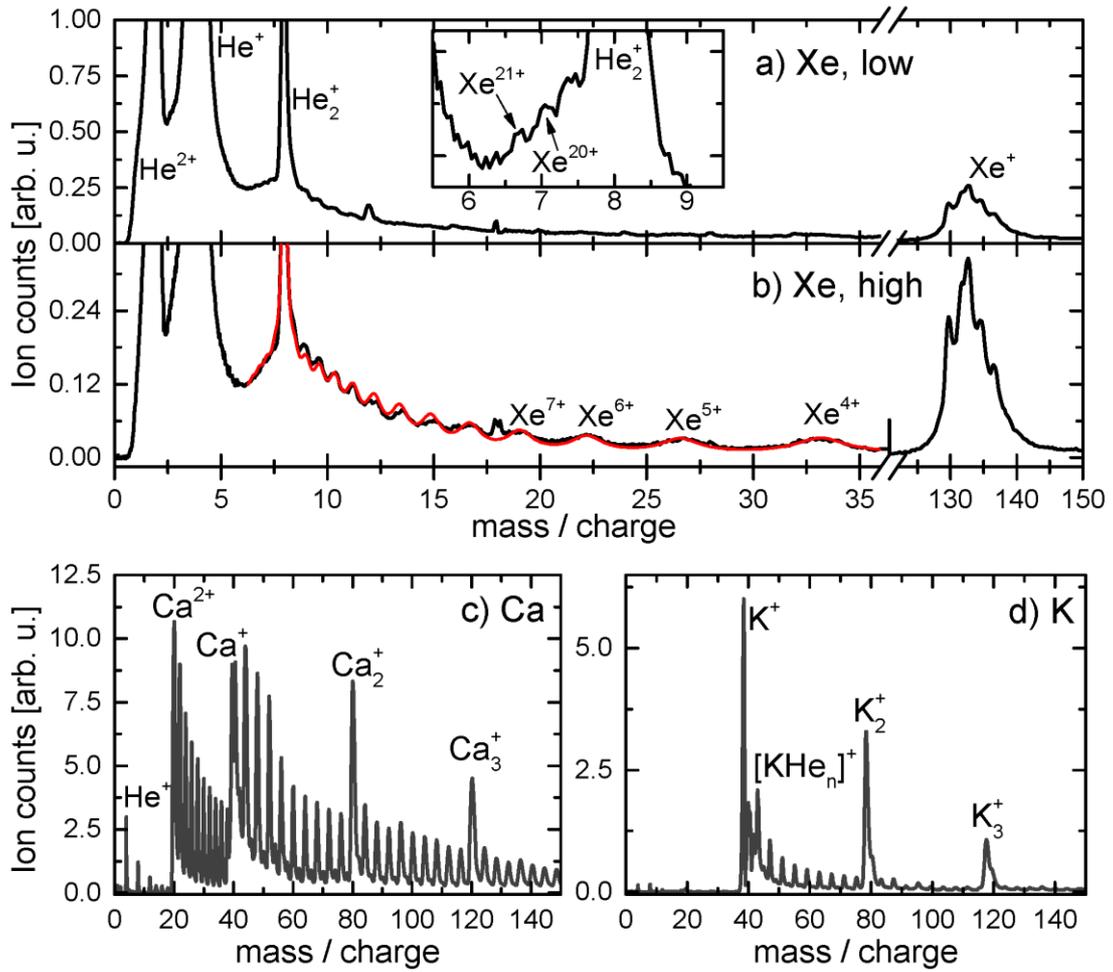

Figure 9. Experimental time-of-flight spectra recorded at a laser intensity $I = 10^{15}$ Wcm$^{-2}$ for He droplets doped with various dopants; a) Average number of dopant atoms $\langle N_{Xe} \rangle \approx 13$; b) $\langle n_{Xe} \rangle \approx 19$; c) $\langle n_{Ca} \rangle \approx 3$; d) $\langle n_K \rangle \approx 10$. The red line in b) exemplifies the nonlinear fit procedure applied for inferring the contributions of different charge states.

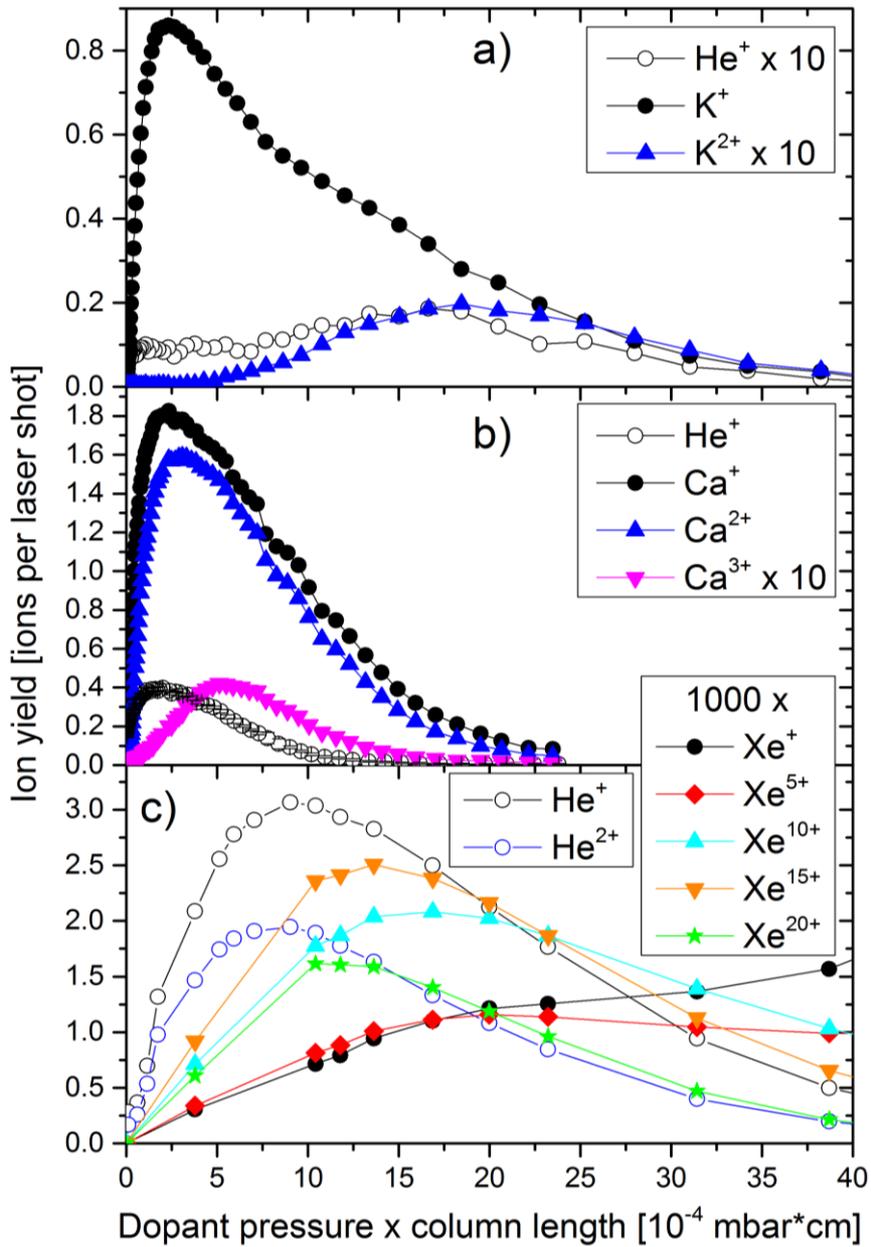

Figure 10. Measured dependence of He and dopant ion yields as a function of the vapor pressure of K, Ca, and Xe dopants multiplied by the length of the doping region (1 cm vapor cell for K, Ca; 35 cm vacuum chamber for Xe).

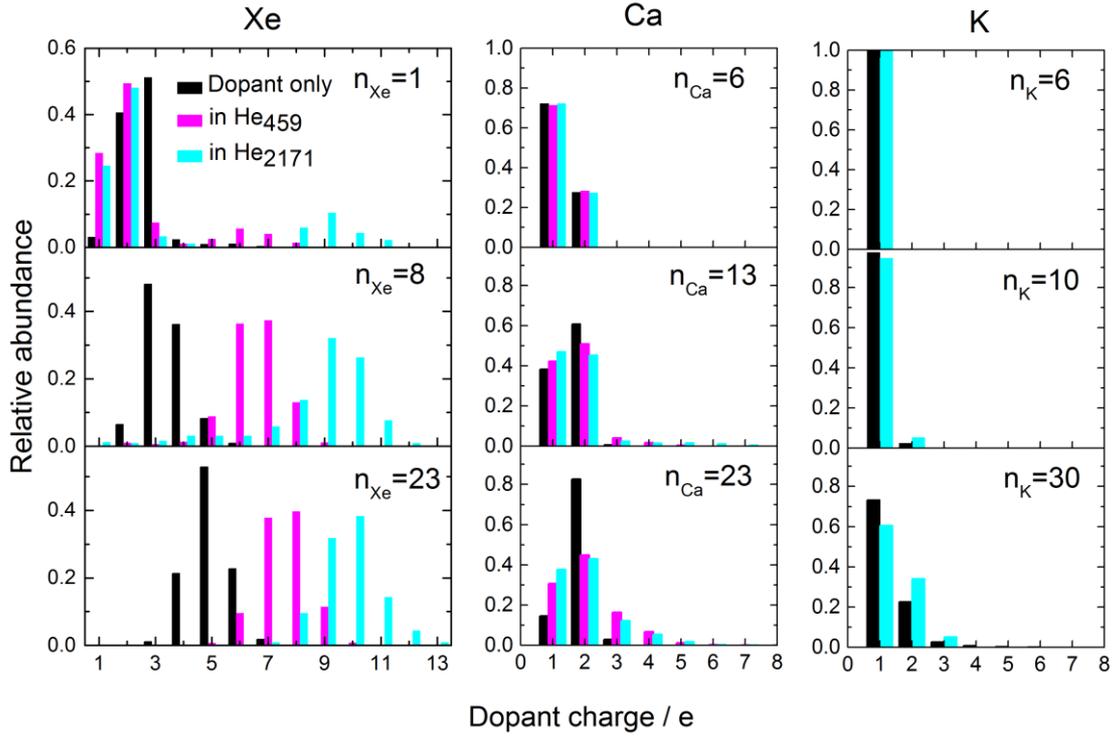

Figure 11. Simulated focally averaged dopant charge state distributions for Xe (left column), Ca (center column), and K dopants (right column). The peak laser intensity is $I = 5 \times 10^{15}$ Wcm$^{-2}$. Magenta and cyan bars are for dopants embedded or attached to He droplets of sizes $N$=459 and 2171, respectively, black bars are for pure dopant clusters.

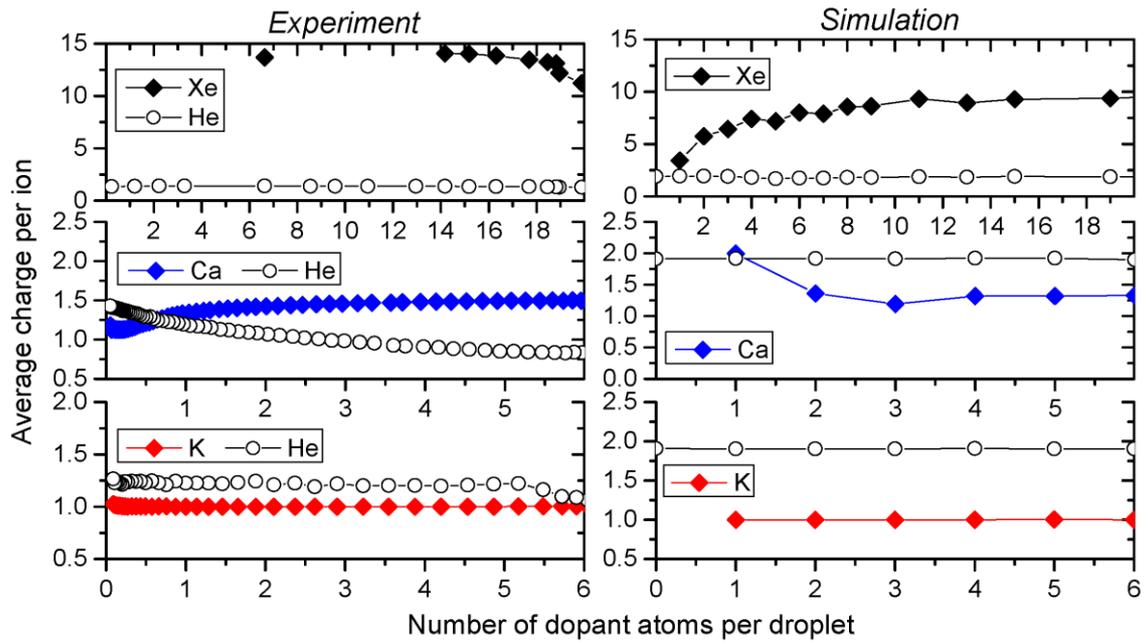

Figure 12. Experimental (left) and simulated (right) average He and dopant ion charge states as a function of the dopant size for all three dopant species (K, Ca, Xe). The experimental average size of the undoped He droplets is $\langle N \rangle = 5000$ atoms but is considerably reduced upon doping depending on the dopant element and dopand cluster size. For the simulations a fixed value of $N = 2171$ is chosen. The simulation results are averaged over the focus volume.